\documentclass[12pt,a4paper]{article}
\usepackage{amssymb}

\usepackage{graphicx}
\usepackage{amsmath}
\usepackage{makeidx}
\usepackage{indentfirst}

\newcounter{resultnum}[section]

\newcounter{conclusionnum}[section]

\newcounter{conditionnum}[section]

\newcounter{conjecturenum}[section]

\newcounter{examplenum}[section]

\newcounter{exercisenum}[section]

\newcounter{lemmanum}[section]

\newcounter{notationnum}[section]

\newcounter{theoremnum}[section]

\newcounter{definitionnum}[section]

\newcounter{corollarynum}[section]

\newcounter{remarknum}[section]

\newcounter{propositionnum}[section]

\newcounter{acknowledgementnum}[section]

\newcounter{algorithmnum}[section]

\newcounter{axiomnum}[section]

\newcounter{casenum}[section]

\newcounter{claimnum}[section]

\newcounter{summarynum}[section]

\newcounter{problemnum}[section]

\begin{document}

\title{Nonholonomic Ricci Flows and\\ Running Cosmological Constant:\\
I. 4D Taub--NUT Metrics }
\author{ Sergiu I. Vacaru\thanks{sergiu$\_$vacaru@yahoo.com,
  svacaru@fields.utoronto.ca } \\
 {\small \it The Fields Institute for Research in Mathematical Science}\\
 {\small \it 222 College Street, Second Floor}\\
 {\small \it Toronto, Ontario M5T 3J1, Canada}
\and
Mihai Visinescu\thanks{mvisin@theory.nipne.ro}\\
{\small \it  Department of Theoretical Physics,}\\
{\small \it National Institute for Physics and Nuclear Engineering} \\
{\small \it P.O. Box M.G.-6, Magurele, Bucharest, Romania}}

\date{September 20, 2006}
\maketitle

\begin{abstract}
In this work we construct and analyze exact solutions describing Ricci flows
and nonholonomic deformations of four dimensional (4D) Taub-NUT
spacetimes. It is outlined a new geometric techniques of constructing Ricci
flow solutions. Some conceptual issues on spacetimes provided with generic
off--diagonal metrics and associated nonlinear connection structures are
analyzed. The limit from gravity/Ricci flow models with nontrivial torsion
to configurations with the Levi-Civita connection is allowed in some
specific physical circumstances by constraining the class of integral
varieties for the Einstein and Ricci flow equations.

\vskip0.3cm



\end{abstract}



\section{ Introduction}
Hawking  has suggested \cite{haw} that the Euclidean Taub-NUT metric
might give rise to the gravitational analogue of the Yang-Mills
instanton. Also, in the long-distant limit, neglecting radiation, the
relative motion of two monopoles is described by the geodesics of this
space \cite{man,atiyah}. The Kaluza-Klein monopole was obtained by
embedding the Taub-NUT gravitational instanton into five-dimensional
Kaluza-Klein theory \cite{sor,gp} (there are various classes of
solitonic/ monopole solutions, in brief termed KK monopoles; see further
developments and reviews of results in pseudo-classical models \cite{vis,dv}).

On the other hand,  the recent experimental data seem to indicate that the
universe does indeed possess a small positive cosmological constant. This
induced a program of researches on generalized solutions in asymptotically
(anti)--de Sitter (AdS) spacetimes (KK--AdS--Taub-NUT solutions \cite%
{onemli,amr,ar} and new cosmological Taub--NUT like solutions \cite{lpp,ms}).

The problem of extending the gravitational monopole solutions to extra
dimensions and/or on spaces with nontrivial cosmological constants is
related to a more general problem of constructing and interpretation of
solutions defining physical objects self--consistently embedded in arbitrary
nontrivial gravitational backgrounds. The usual techniques for generating
new classes of monopole solutions with nonzero cosmological constants is to
add the time like coordinate, to generalize the metric to time dependencies
(for instance, to a cosmological model) and than to perform the
Kaluza--Klein compactification on the fifth dimension (in brief, 5D).

There is a more general approach of constructing exact solutions in gravity
following the so--called 'anholonomic frame method' elaborated and developed
in a series of works (see Refs. \cite{vhep2,vs,vesnc}). The idea is to use
certain classes of nonholnomic (equivalently, anholonomic) deformations of
the frame, metric and connection structures and superpositions of
generalized conformal maps in order to generate a class of off--diagonal
metric ansatz solving exactly the vacuum or nonvacuum Einstein equations.%
\footnote{%
Such ansatz can not be diagonalized by coordinate transforms but can be
effectively diagonalized with respect to certain systems of nonholonomic
local frames with associated nonlinear connection structures, see details in %
\cite{vesnc}. This allows us to apply a well developed geometric techniques
in order to define generalized symmetries and to integrate exactly the
corresponding systems of field equations. For instance, for 5D spacetimes,
such solutions depend on sets of integrating functions on four and three
variables.} The method was considered, for instance, for constructing
locally anisotropic Taub-NUT solutions \cite{vt} and investigating
self--consistent propagations of three dimensional Dirac and/or solitonic
waves in such spacetimes \cite{vp}.

A general nonholonomic transform of a ''primary'' metric (it can be an exact
solution or, for instance, a conformal transform of a known exact solution)
into a generic off--diagonal exact solution does not preserve the properties
of former metric. Nevertheless, if there are satisfied certain smooth limits
to already known solutions, boundary (asymptotic) and deforming symmetry
conditions, one may consider that, for instance, a primary Taub-NUT
configuration became locally anisotropic with polarized constants and/or
imbedded self--consistently in a nontrivial, for instance, solitonic
background. The new classes of solutions can be used for testing various
type of physical theories (string/brane gravity, noncommutative and gauge
gravity, Finsler like generalizations ...) when the physical interactions
are described by generic off--diagonal metrics and nonholonomic constraints;
in general, with nontrivial topological configurations and new type
symmetries (non--Killing ones, for instance, with generalized Lie algebra
symmetries); the extra dimensions are not subjected to the Kaluza Klein
restrictions.

The anholonomic frame method seems to be effective in constructing exact
solutions of the Ricci flow equations \cite{vrf}. Such equations were
introduced by R. Hamilton \cite{ham1} in order to describe the geometric
evolution of a Riemannian manifold $\left( V,g\right) ,$  where by $g$ we
denote the metric, in the direction of its Ricci tensor $Ric(g),$ see
reviews of results and applications in \cite{ham2,aubin,cao}. There is a
vast potential use of this geometric approach in theoretical physics and
cosmology \cite{per,bakas,geg,crvis,dm,wg,hw}.

It is known that there is a unique solution for the Ricci flow with bounded
curvature on a complete noncompact manifold \cite{chen}. A very important
task is to construct exact solutions for the Ricci flow equations and to
investigate their physical implications. The purpose of this work is
two--fold. The first one is to elaborate a general method of constructing
exact Ricci flow solutions for certain classes of off--diagonally deformed
metric ansatz, both for connections with trivial and nontrivial torsion. The
other is to construct explicit examples of such 4D exact solutions, for
Taub-NUT like metrics, to analyze their physical properties and to show that
they may possess nontrivial limits to the Einstein spaces.

The structure of the paper is as follows: In section 2, we define a new
geometric approach to constructing exact solutions for Ricci flows of
generic off--diagonal metrics. The approach is based on the so--called
anholonomic frame method with associated nonlinear connection structure. We
construct a general class of integral varieties of Ricci flow equations and
define the constrains for the Levi-Civita configurations. In Section 3, we
apply the formalism to 4D Taub--NUT configurations and their off--diagonal
Ricci flows. The last section is devoted to conclusions and discussion. In
Appendix A, we outline the necessary material from the geometry of nonlinear
connections and related nonholonomic deformations.

\section{Nonholonomic Ricci flows}

In this section we introduce an off--diagonal ansatz for metrics depending
on three coordinates and consider the anholonomic frame method of
constructing exact solutions for the system of equations defining Ricci
flows of metrics subjected to nonholonomic constraints.

\subsection{Geometric preliminaries}

The normalized Ricci flows \cite{ham1,ham2,aubin,cao,per,bakas}, with
respect to a coordinate base $\partial _{\underline{\alpha }}=\partial
/\partial u^{\underline{\alpha }},$ are defined by the equations
\begin{equation}
\frac{\partial }{\partial \tau }g_{\underline{\alpha }\underline{\beta }%
}=-2R_{\underline{\alpha }\underline{\beta }}+\frac{2r}{5}g_{\underline{%
\alpha }\underline{\beta }},  \label{feq}
\end{equation}%
where $R_{\underline{\alpha }\underline{\beta }}$ is the Ricci tensor of a
metric $g_{\underline{\alpha }\underline{\beta }}$ and corresponding
Levi--Civita connection\footnote{By definition this connection is torsionless and
metric
compatible.} and the normalizing factor $r=\int RdV/dV$ is introduced in
order to preserve the volume $V.$ In this work we shall consider flows of
generic off--diagonal four dimensional (4D) metrics
\begin{equation}
\mathbf{g}=g_{\underline{\alpha }\underline{\beta }}(u)du^{\underline{\alpha
}}\otimes du^{\underline{\beta }},  \label{metr}
\end{equation}%
where
\begin{equation*}
g_{\underline{\alpha }\underline{\beta }}=\left[
\begin{array}{cc}
g_{ij}+N_{i}^{a}N_{j}^{b}h_{ab} & N_{j}^{e}h_{ae} \\
N_{i}^{e}h_{be} & h_{ab}%
\end{array}%
\right] ,
\end{equation*}%
with the indices of type $\underline{\alpha },\underline{\beta }%
=(i,a),(j,b)...$ running the values $i,j,...=1,2$ and $a,b,...=3,4$ (we
shall omit underlying of indices for the components of arbitrary basis or
even with respect to coordinate basis if that will not result in
ambiguities) and local coordinates labelled in the form $u=(x,y)=\{u^{\alpha
}=(x^{i},y^{a})\}.$

Applying the frame transforms
\begin{equation*}
\mathbf{e}_{\alpha }=\mathbf{e}_{\alpha }^{ \underline{\alpha }}\partial _{%
\underline{\alpha }}\mbox{ and }\mathbf{c}^{\alpha }=\mathbf{e}_{\
\underline{\alpha }}^{\alpha  }du^{\underline{\alpha }}
\end{equation*}%
for $\mathbf{e}_{\alpha }\rfloor \mathbf{c}^{\beta }=\delta _{\alpha
}^{\beta },$ where ''$\rfloor $'' denotes the inner product and $\delta
_{\alpha }^{\beta }$ is the Kronecker symbol, with
\begin{equation}
\mathbf{e}_{\alpha }^{\ \underline{\alpha }}=\left[
\begin{array}{cc}
\delta _{i}^{\ \underline{i}} & N_{i}^{b}(u)\delta _{b}^{\ \underline{a}} \\
0 & \delta _{b}^{\ \underline{a}}%
\end{array}%
\right] \mbox{ and }\mathbf{e}_{\ \underline{\alpha }}^{\alpha \ }=\left[
\begin{array}{cc}
\delta _{\ \underline{i}}^{i} & -N_{k}^{b}(u)\delta _{\ \underline{i}}^{k}
\\
0 & \delta _{\ \underline{a}}^{a\ }%
\end{array}%
\right] ,  \label{nhft}
\end{equation}%
we represent the metric (\ref{metr}) in an effectively diagonalized $(n+m)$%
--distingu\-ish\-ed form, see the metric (\ref{m2}) in Appendix,
\begin{eqnarray}
\mathbf{g} &=&g_{\alpha }(u)\ \mathbf{c}^{\alpha }\otimes \mathbf{c}^{\alpha
}=g_{i}(u)\ b^{i}\otimes b^{i}+h_{a}(u)\ b^{a}\otimes b^{a},  \notag \\
\mathbf{c}^{\alpha } &=&(b^{i}=dx^{i},b^{a}=dy^{a}+N_{i}^{a}(u)dx^{i}).
\label{m2a}
\end{eqnarray}%
Such metrics and frame transforms are considered in the geometry of
nonholonomic manifolds with associated N--connection structure defined by
the set $\mathbf{N}=\{N_{k}^{b}\}$ stating a nonintegrable distribution on a
4D manifold $\mathbf{V}.$ In Appendix A, we outline the geometry of such
spaces. In this paper, we shall consider classes of solutions with ansatz of
type (\ref{m2a}) (and (\ref{m1}) or (\ref{m2})), when
\begin{equation}
g_{i}=g_{i}(x^{k}),h_{a}=h_{a}(x^{k},v),
N_{i}^{3}=w_{i}(x^{k},v),N_{i}^{4}=n_{i}(x^{k},v),  \label{m2ac}
\end{equation}%
for $y^{3}=v$ being the so--called ''anisotropic'' coordinate. Such metrics
are very general off--diagonal ones, with the coefficients depending on 2
and 3 coordinates but not depending on the coordinate $y^{4}.$

To consider flows of metrics related both to the Einstein and
string gravity (in the last case there is a nontrivial antisymmetric torsion
field), it is convenient to work with the so--called canonical distinguished
connection (in brief, d--connection) $\widehat{\mathbf{D}}=\{\widehat{%
\mathbf{\Gamma }}_{\ \alpha \beta }^{\gamma }\}$ which is metric compatible
but with nontrivial torsion (see formulas (\ref{candcon}) and related
discussions). Imposing certain restrictions on the coefficients $N_{k}^{b},$
see (\ref{fols}), (\ref{coef0}) and (\ref{cond3}), we can satisfy the
conditions that the coefficients of the canonical d--connection of the
Levi--Civita $\bigtriangledown =\{\ _{\shortmid }\Gamma _{\beta \gamma }^{\alpha
}\}$ are defined by the same nontrivial values $\widehat{\mathbf{\Gamma }}%
_{\ \alpha \beta }^{\gamma }=\ _{\shortmid }\Gamma _{\ \alpha \beta
}^{\gamma }$ with respect to N--adapted basis (\ref{b1a}) and (\ref{cb1a}).

The Ricci flow equations (\ref{feq}) can be written for the Ricci tensor of
the canonical d--connection (\ref{dricci}) and metric (\ref{m2a}), as it was
considered in Ref. \cite{vrf},
\begin{eqnarray}
\frac{\partial }{\partial \tau }g_{ij} &=&-2R_{ij}+2\lambda g_{ij}-h_{cd}%
\frac{\partial }{\partial \tau }(N_{i}^{c}N_{j}^{d}),  \label{1eq} \\
\frac{\partial }{\partial \tau }h_{ab} &=&-2S_{ab}+2\lambda h_{ab},\
\label{2eq} \\
R_{\alpha \beta } &=&0\mbox{ and }g_{\alpha \beta }=0\mbox{ for }\alpha \neq
\beta ,\text{ }  \notag
\end{eqnarray}%
where $\lambda =r/5,$ $y^{3}=v$ and $\tau $ can be, for instance, the time
like  coordinate, $\tau =t,$ or any parameter or extra dimension
coordinate. The equations (\ref{1eq}) and (\ref{2eq}) are just the
nonholonomic frame transform with the matrices (\ref{nhft}) of the equations
(\ref{feq}). The aim of this section is to show how the anholonomic frame
method developed in \cite{vrf} (for Ricci flows) and in \cite%
{vhep2,vs,vesnc,vt,vp} (for off--diagonal exact solutions) can be used for
constructing exact solutions of the system of Ricci flow equations (\ref{1eq}%
) and (\ref{2eq}) describing nonholnomic deformations of 3D and 4D Taub-NUT
solutions.

\subsection{Ricci flow equations for off--diagonal metric ansatz}

We consider a primary (pseudo) Riemannian metric
\begin{eqnarray}
\mathbf{\check{g}} &=&\check{g}_{1}(x^{k},v,y^{4})(dx^{1})^{2}+\check{g}%
_{2}(x^{k},v,y^{4})(dx^{2})^{2}  \notag \\
&&+\check{h}_{3}(x^{k},v,y^{4})(\check{b}^{3})^{2}+\check{h}%
_{4}(x^{k},v,y^{4})(\check{b}^{4})^{2},  \notag \\
\check{b}^{3} &=&dv+\check{w}_{i}(x^{k},v,y^{4})\ dx^{i},\ \check{b}%
^{4}=dy^{4}+\check{n}_{i}(x^{k},v,y^{4})\ dx^{i},  \label{m1a}
\end{eqnarray}%
which is a particular case of metric (\ref{m1}) with N--connection
coefficients $\check{N}_{i}^{3}=\check{w}_{i}(x^{k},v,y^{4})$ and $\check{N}%
_{i}^{4}=\check{n}_{i}(x^{k},v,y^{4})$ for $y^{3}=v$ considered to be a
nonholonomically constrained coordinate. The metric
may be an exact solution of the Einstein equations, or any conformal
transform of a such one (which is not an exact solution).\footnote{%
In this work, we shall consider the primary metrics to be related to certain
Taub-NUT like solutions, or some their conformal transforms and/or trivial
embedding / compactification in extra/lower dimension spaces.} An
anholonomic transform, $\mathbf{\check{N}\rightarrow N}$ and $\mathbf{\check{%
g}=}(\check{g},\check{h})\rightarrow \mathbf{g=}(g,h),$ can defined by
formulas (\ref{polf}),
\begin{equation*}
g_{i}=\eta _{i}(x^{k},v,y^{4})\check{g}_{i}, h_{a}=\eta _{a}(x^{k},v,y^{4})%
\check{h}_{a}, N_{i}^{a}=\eta _{i}^{a}(x^{k},v,y^{4})\check{N}_{i}^{a}
\end{equation*}%
when the polarizations $\eta _{\alpha }$ and $\eta _{i}^{a}$ are chosen in a
form that the ''target'' metric $\mathbf{g}$  has coefficients of type (\ref%
{m2ac}), i.e. it is parametrized in the form
\begin{eqnarray}
\mathbf{g}
&=&g_{1}(x^{k})(dx^{1})^{2}+g_{2}(x^{k})(dx^{2})^{2}+h_{3}(x^{k},v)(b^{3})^{2}
+h_{4}(x^{k},v)(b^{4})^{2},
\notag \\
b^{3} &=&dv+w_{i}(x^{k},v)\ dx^{i},\ b^{4}=dy^{4}+n_{i}(x^{k},v)\ dx^{i},
\label{m2b}
\end{eqnarray}%
defining an exact solution for the nonholonomic Ricci flow equations (\ref%
{1eq}) and (\ref{2eq}), when $\tau $ is one of the coordinates $x^{k},$ or $%
v.$

The nontrivial components of the Ricci tensor $R_{\alpha \beta }$ (\ref%
{dricci}) (see details of a similar calculus in Ref. \cite{vesnc}) are
\begin{eqnarray}
R_{\ 1}^{1} &=&R_{\ 2}^{2}=\frac{1}{2g_{1}g_{2}}\left[ \frac{g_{1}^{\bullet
}g_{2}^{\bullet }}{2g_{1}}+\frac{(g_{2}^{\bullet })^{2}}{2g_{2}}%
-g_{2}^{\bullet \bullet }+\frac{g_{1}^{^{\prime }}g_{2}^{^{\prime }}}{2g_{1}}%
+\frac{(g_{1}^{^{\prime }})^{2}}{2g_{2}}-g_{1}^{^{\prime \prime }}\right] ,
\label{rth} \\
S_{\ 3}^{3} &=&S_{\ 4}^{4}=\frac{1}{2h_{3}h_{4}}\left[ -h_{4}^{\ast \ast }+%
\frac{\left( h_{4}^{\ast }\right) ^{2}}{2h_{4}}+\frac{h_{4}^{\ast
}h_{3}^{\ast }}{2h_{4}}\right] ,  \label{rtv} \\
R_{3i} &=&-\frac{1}{2h_{4}}\left( w_{i}\beta +\alpha _{i}\right) ,
\label{3eq} \\
R_{4i} &=&-\frac{h_{4}}{2h_{3}}\left( n_{i}^{\ast \ast }+\gamma n_{i}^{\ast
}\right)   \label{4eq}
\end{eqnarray}%
where
\begin{eqnarray}
\alpha _{i} &=&\partial _{i}h_{4}^{\ast }-h_{4}^{\ast }\partial _{i}\ln
\sqrt{\left| h_{3}h_{4}\right| },\ \beta =h_{4}^{\ast \ast }-h_{4}^{\ast
}\left( \ln \sqrt{\left| h_{3}h_{4}\right| }\right) ^{\ast },  \label{aux1}
\\
\gamma  &=&3h_{4}^{\ast }/2h_{4}-h_{3}^{\ast }/h_{3},\mbox{ for }h_{3}^{\ast
}\neq 0,h_{4}^{\ast }\neq 0,  \notag
\end{eqnarray}%
defined by $h_{3}$ and $h_{4}$ as solutions of equations (\ref{2eq}). In the
above presented formulas, it was convenient to write the partial derivatives
in the form $a^{\bullet }=\partial a/\partial x^{1},a^{^{\prime }}=\partial
a/\partial x^{2}$ and $a^{\ast }=\partial a/\partial v.$

We consider a general method of constructing solutions of the Ricci flows
equations related to the so--called Einstein spaces with nonhomogeneously
polarized cosmological constant, when
\begin{eqnarray}
R_{\ j}^{i} &=&\lambda _{\lbrack h]}(x^{k})\ \delta _{\ j}^{i},\ S_{\
b}^{a}=\lambda _{\lbrack v]}(x^{k},v)\ \delta _{\ b}^{a},  \notag \\
R_{\alpha \beta } &=&0\mbox{ and }g_{\alpha \beta }=0\mbox{ for }\alpha \neq
\beta ,\text{ }  \notag
\end{eqnarray}%
when $\lambda _{\lbrack h]}$ and $\lambda _{\lbrack v]}$ are induced by
certain string gravity ansatz (\ref{ansh}), or matter field contributions.
In particular case, we can fix $\lambda _{\lbrack h]}=\lambda _{\lbrack v]}$
and consider off--diagonal metrics of the usual Einstein spaces with
cosmological constant.

The nonholonomic Ricci flows equations (\ref{1eq}) and (\ref{2eq}) for the
Einstein spaces with nonhomogeneous cosmological constant defined by ansatz
of type (\ref{m2b}) transform into the following system of partial
differential equations consisting from two subsets of equations: The first
subset of equations consists from those generated by the Einstein equations
for the off--diagonal metric,
\begin{eqnarray}
\frac{g_{1}^{\bullet }g_{2}^{\bullet }}{2g_{1}}+\frac{(g_{2}^{\bullet })^{2}%
}{2g_{2}}-g_{2}^{\bullet \bullet }+\frac{g_{1}^{^{\prime }}g_{2}^{^{\prime }}%
}{2g_{1}}+\frac{(g_{1}^{^{\prime }})^{2}}{2g_{2}}-g_{1}^{^{\prime \prime }}
&=&2g_{1}g_{2}\ \lambda _{\lbrack h]}(x^{k}),  \label{ee1} \\
-h_{4}^{\ast \ast }+\frac{\left( h_{4}^{\ast }\right) ^{2}}{2h_{4}}+\frac{%
h_{4}^{\ast }h_{3}^{\ast }}{2h_{4}} &=&2h_{3}h_{4}\ \lambda _{\lbrack
v]}(x^{k},v),  \label{ee2} \\
w_{i}\beta +\alpha _{i} &=&0,  \label{ee3} \\
n_{i}^{\ast \ast }+\gamma n_{i}^{\ast } &=&0.  \label{ee4}
\end{eqnarray}
The second subset of equations is formed just by those describing flows of
the diagonal, $g_{ij}=diag[g_{1},g_{2}]$ and $h_{ab}=diag[h_{3},h_{4}],$ and
off diagonal, $w_{i}$ and $n_{i},$ metric coefficients,
\begin{eqnarray}
\frac{\partial }{\partial \tau }g_{ij} &=&2\lambda _{\lbrack h]}(x^{k})\
g_{ij}-h_{3}\frac{\partial }{\partial \tau }\left( w_{i}w_{j}\right) -h_{4}%
\frac{\partial }{\partial \tau }\left( n_{i}n_{j}\right) ,  \label{rfe1} \\
\frac{\partial }{\partial \tau }h_{a} &=&2\lambda _{\lbrack v]}(x^{k},v)\
h_{a}.  \label{rfe2}
\end{eqnarray}%
The aim of the next section is to show how we can integrate the equations (%
\ref{ee1})--(\ref{rfe2}) in a quite general form.

\subsection{Integral varieties of solutions of Ricci flow equations}

We emphasize that the system of equations (\ref{ee1})--(\ref{ee4}) was
derived and solved in general form for a number of 4D and 5D metric ansatz
of type (\ref{m2b}), or (\ref{m2}), in Refs. \cite{vhep2,vs,vesnc,vt,vp} for
various models of gravity theory. The idea of work \cite{vrf} was to use the
former method and some integral varieties of those solutions in order to
subject the metric and N--connection coefficients additionally to the
conditions (\ref{rfe1}) and (\ref{rfe2}) and generate Ricci flows of
off--diagonal metrics. Here, we briefly outline the method of constructing
such general solutions.

We begin with equation (\ref{ee1}) for the metric coefficients $%
g_{i}(x^{k})$ on a 2D subspace. By a corresponding coordinate transform $x^{%
\widetilde{i}}\rightarrow x^{\widetilde{i}}(x^{i}),$ such metrics can be
always diagonalized and represented in conformally flat form,%
\begin{equation*}
g_{i}(x^{k})(dx^{i})^{2}=e^{\psi (x^{\widetilde{i}})}\left[ \epsilon
_{1}(dx^{\widetilde{1}})^{2}+\epsilon _{2}(dx^{\widetilde{2}})^{2}\right],
\end{equation*}%
where the values $\epsilon _{i}=\pm 1$ depend on chosen signature. The
equation (\ref{ee1}) transforms into
\begin{equation}
\epsilon _{1}\psi ^{\bullet \bullet }+\epsilon _{2}\psi ^{^{\prime \prime
}}=2\lambda _{\lbrack h]}(x^{k}).  \label{ee1a}
\end{equation}%
Such equations and their equivalent 2D coordinate transform can be written
in three alternative ways convenient for different types of nonholonomic
deformations of metrics. For instance, we can prescribe that $g_{1}=g_{2}$
and write the equation for $\psi =\ln |g_{1}|=\ln |g_{2}|.$ Alternatively,
we can suppose that $g_{1}^{\prime }=0$ for a given $g_{1}(x^{1})$ (or $%
g_{2}^{\bullet }=0,$ for a prescribed $g_{2}(x^{2}))$ and get from (\ref{ee1}%
) the equation%
\begin{equation*}
\frac{g_{1}^{\bullet }g_{2}^{\bullet }}{2g_{1}}+\frac{(g_{2}^{\bullet })^{2}%
}{2g_{2}}-g_{2}^{\bullet \bullet }=2g_{1}g_{2} \lambda _{\lbrack h]}(x^{k})
\end{equation*}%
(or%
\begin{equation*}
\frac{g_{1}^{^{\prime }}g_{2}^{^{\prime }}}{2g_{1}}+\frac{(g_{1}^{^{\prime
}})^{2}}{2g_{2}}-g_{1}^{^{\prime \prime }}=2g_{1}g_{2} \lambda _{\lbrack
h]}(x^{k}),
\end{equation*}%
in the inverse case). In general, we can prescribe that, for instance, $%
g_{1}(x^{k})$ is defined by any solution of the 2D
Laplace/D'Alambert/solitonic equation and try to define $g_{2}(x^{k})$
constrained to be satisfied one of the above equations related to (\ref{ee1}%
). We conclude that such 2D equations can be solved always in explicit or
non explicit form.

Equation (\ref{ee2}) relates two nontrivial v--coefficients of the
metric coefficients $h_{3}(x^{k},v)$ and $h_{4}(x^{k},v)$ depending on three
coordinates but with partial derivatives only on the third (anisotropic)
coordinate. As a matter of principle, we can fix $h_{3}$ (or, inversely, $%
h_{4}$) to describe any physically interesting situation being, for
instance, a solution of the 3D solitonic, or pp--wave equation,  and than
we can try to define $h_{4}$ (inversely, $h_{3}$) in order to get a solution
of (\ref{ee2}). Here we note that it is possible to solve such equations for
any $\lambda _{\lbrack v]}(x^{k},v),$ in general form, if $h_{4}^{\ast }\neq
0$ (for $h_{4}^{\ast }=0,$ there are nontrivial solutions only if $\lambda
_{\lbrack v]}=0).$ Introducing the function
\begin{equation}
\phi (x^{i},v)=\ln \left| h_{4}^{\ast }/\sqrt{|h_{3}h_{4}|}\right| ,
\label{afa1}
\end{equation}%
we write that equation in the form%
\begin{equation}
\left( \sqrt{|h_{3}h_{4}|}\right) ^{-1}\left( e^{\phi }\right) ^{\ast
}=-2\lambda _{\lbrack v]}.  \label{afa2}
\end{equation}%
Using (\ref{afa1}), we express $\sqrt{|h_{3}h_{4}|}$ as a function of $\phi $
and $h_{4}^{\ast }$ and obtain%
\begin{equation}
|h_{4}^{\ast }|=-(e^{\phi })^{\ast }/4\lambda _{\lbrack v]}  \label{fa1}
\end{equation}%
which can be integrated in general form,
\begin{equation}
h_{4}=h_{4[0]}(x^{i})-\frac{1}{4}\int dv\ \frac{\left[ e^{2\phi (x^{i},v)}%
\right] ^{\ast }}{\lambda _{\lbrack v]}(x^{i},v)},  \label{solh4}
\end{equation}%
where $h_{4[0]}(x^{i})$ is the integration function. Having defined $h_{4}$
and using again (\ref{afa1}), we can express $h_{3}$ via $h_{4}$ and $\phi ,$
\begin{equation}
|h_{3}|=4e^{-2\phi (x^{i},v)}\left[ \left( \sqrt{|h_{4}|}\right) ^{\ast }%
\right] ^{2}.  \label{solh5}
\end{equation}%
The conclusion is that prescribing any two functions $\phi (x^{i},v)$ and $%
\lambda _{\lbrack v]}(x^{i},$ $v)$ we can always find the corresponding
metric coefficients $h_{3}$ and $h_{4}$ solving (\ref{ee2}). Following (\ref%
{solh5}), it is convenient to represent such solutions in the form%
\begin{eqnarray*}
h_{4} &=&\epsilon _{4}\left[ b(x^{i},v)-b_{0}(x^{i})\right] ^{2} \\
h_{3} &=&4\epsilon _{3}e^{-2\phi (x^{i},v)}\left[ b^{\ast }(x^{i},v)\right]
^{2}
\end{eqnarray*}%
where $\epsilon _{a}=\pm 1$ depending on fixed signature, $b_{0}(x^{i})$ and
$\phi (x^{i},v)$ can be arbitrary functions and $b(x^{i},v)$ is any function
when $b^{\ast }$ is related to $\phi $ and $\lambda _{\lbrack v]}$ as stated
by the formula (\ref{fa1}). Finally, we note that if $\lambda _{\lbrack
v]}=0,$ we can relate $h_{3}$ and $h_{4}$ solving (\ref{afa2}) as $\left(
e^{\phi }\right) ^{\ast }=0.$

For any couples $h_{3}$ and $h_{4}$ related by (\ref{ee2}), we can compute
the values $\alpha _{i},\beta $ and $\gamma $ (\ref{aux1}). This allows us
to define the off--diagonal metric (N--connection) coefficients $w_{i}$
solving (\ref{ee3}) as algebraic equations,%
\begin{equation}
w_{i}=-\alpha _{i}/\beta =-\partial _{i}\phi /\phi ^{\ast }.  \label{see3}
\end{equation}%
We emphasize, that for the vacuum Einstein equations one can be solutions of
(\ref{ee2}) resulting in $\alpha _{i}=\beta =0.$ In such cases, $w_{i}$ can
be arbitrary functions on variables $(x^{i},v)$ with finite values for
derivatives in the limits $\alpha _{i},\beta \rightarrow 0$ eliminating the
''ill--defined'' situation $w_{i}\rightarrow 0/0.$ For the Ricci flow
equations with nonzero values of $\lambda _{\lbrack v]},$ such difficulties
do not arise. The second subset of N--connection (off--diagonal metric)
coefficients $n_{i}$ can be computed by integrating two times on variable $v$
in (\ref{ee4}), for given values $h_{3}$ and $h_{4}.$ One obtains%
\begin{equation}
n_{k}=n_{k[1]}(x^{i})+n_{k[2]}(x^{i})\ \hat{n}_{k}(x^{i},v),  \label{see4}
\end{equation}%
where
\begin{eqnarray*}
\hat{n}_{k}(x^{i},v) &=&\int h_{3}(\sqrt{|h_{4}|})^{-3}dv, h_{4}^{\ast
}\neq 0;  \\
&=&\int h_{3}dv, h_{4}^{\ast }=0; \\
&=&\int (\sqrt{|h_{4}|})^{-3}dv, h_{3}^{\ast }=0,
\end{eqnarray*}%
and $n_{k[1]}(x^{i})$ and $n_{k[2]}(x^{i})$ are integration functions.

We conclude that any solution $\left( h_{3},h_{4}\right) $ of the equation (%
\ref{ee2}) with $h_{4}^{\ast }\neq 0$ and non vanishing $\lambda _{\lbrack
v]} $ generates the solutions (\ref{see3}) and (\ref{see4}), respectively,
of equations (\ref{ee3}) and (\ref{ee4}). Such solutions (of the Einstein
equations) are defined by the mentioned classes of integration functions and
prescribed values for $b(x^{i},v)$ and $\psi (x^{i}).$ Further restrictions
on $(g_{1},g_{2})$ and $\left( h_{3},h_{4}\right) $ are necessary in order
to satisfy the equations (\ref{rfe1}) and (\ref{rfe2}) relating flows of the
metric and N--connection coefficients in a compatible manner. It is not
possible to solve in a quite general form such equations, but in the next
section we shall give certain examples of such solutions defining flows of
the Taub-NUT like metrics.

\subsection{Extracting solutions for the Levi-Civita connection}

The method outlined in the previous section allows us to construct integral
varieties for the Ricci flow equations (\ref{ee1})--(\ref{rfe2}) derived for
the canonical d--connection with nontrivial torsion, see formulas (\ref%
{candcon}) and (\ref{dtors}) in Appendix. We can restrict such integral
varieties (constraining the off--diagonal metric, equivalently,
N--connection coefficients $w_{i}$ and $n_{i}$ and related integration
functions) in order to generate solutions for the Levi-Civita connection.
The conditions $_{\shortmid }\Gamma _{\beta \gamma }^{\alpha }=\widehat{%
\mathbf{\Gamma }}_{\ \alpha \beta }^{\gamma }$ (i.e. the coefficients of the
Levi-Civita connection are equal to the coefficients of the canonical
d--connection, both classes of coefficients being computed with respect to
the N--adapted bases (\ref{b1a}) and (\ref{cb1a})) hold true if there are
satisfied the equations (\ref{fols}), (\ref{coef0}) and (\ref{cond3}).%
\footnote{%
We emphasize that connections on manifolds are not defined as tensor objects. If the
coefficients of two different connections are equal with respect to one frame,
they can be very different with respect to other frames.}

Introducing the coefficients for the ansatz (\ref{m2b}), we can see that the
constrains (\ref{coef0}) are trivially satisfied and the equations (\ref%
{cond3}) are written in the form%
\begin{eqnarray}
\frac{\partial h_{3}}{\partial x^{k}}-w_{k}h_{3}^{\ast }-2w_{k}^{\ast }h_{3}
&=&0,  \label{cond3a} \\
\frac{\partial h_{4}}{\partial x^{k}}-w_{k}h_{4}^{\ast } &=&0,
\label{cond3b} \\
n_{k}^{\ast }h_{4} &=&0.  \label{cond3c}
\end{eqnarray}%
The relations (\ref{cond3a}) and (\ref{cond3b}) are equivalent for the
general solutions $h_{3},$ see (\ref{solh5}), $h_{4},$ see (\ref{solh4}) and
$w_{i},$ see (\ref{see3}), generated by a function $\phi (x^{i},v)$ (\ref%
{afa1}) if $\phi \rightarrow \phi -\ln 2,$ when
\begin{equation*}
\phi =\ln |\left( \sqrt{|h_{4}|}\right) ^{\ast }|-\ln |\left( \sqrt{|h_{3}|}%
\right) |
\end{equation*}%
and
\begin{equation*}
w_{k}=(h_{4}^{\ast })^{-1}\frac{\partial h_{4}}{\partial x^{k}}=-(\phi
^{\ast })^{-1}\frac{\partial \phi }{\partial x^{k}},
\end{equation*}%
where $\phi =const$ is possible only for the vacuum Einstein solutions. The
condition (\ref{cond3c}) for $h_{4}\neq 0$ constrains $n_{k}^{\ast }=0$
which holds true if we put the integration functions $n_{k[2]}=0$ in (\ref%
{see4}), when $n_{k}=n_{k[1]}(x^{i}).$ These values of $w_{k}$ and $n_{k}$
have to be constrained one more again in order to solve the equations (\ref%
{fols}), which for our ansatz are of type
\begin{eqnarray}
w_{1}^{\prime }-w_{2}^{\bullet }+w_{2}w_{1}^{\ast }-w_{1}w_{2}^{\ast } &=&0,
\label{aux31a} \\
n_{1}^{\prime }-n_{2}^{\bullet } &=&0,  \label{aux31b}
\end{eqnarray}%
stating integrable (pseudo) Riemannian foliations. We have to take such
integration functions when (\ref{aux31b}) is satisfied from the very
beginning for some two functions $n_{k[1]}(x^{i})$ depending on two
variables. In a particular case, we can consider any parametrization of type
$w_{k}=\widehat{w}_{k}(x^{i})q(v)$ for some functions $\widehat{w}%
_{k}(x^{i}) $ and $q(v)$ defining a class of solutions of (\ref{aux31a}).

The final conclusion in this section is that taking any solution of
equations (\ref{ee2}), (\ref{ee3}) and (\ref{ee4}) we can restrict the
integral varieties to integration functions satisfying the conditions (\ref%
{aux31a}) and (\ref{aux31b}) allowing us to extract torsionless
configurations for the Levi-Civita connection.

\section{Nonholonomic Ricci Flows and 4D Taub-NUT Spaces}

The techniques elaborated in previous section can be applied in order to
generate Ricci flow solutions for various classes 4D metrics. In this
section, we examine such configurations derived from a primary Taub-NUT
metric.

We begin with the primary quadratic element
\begin{equation}
d\check{s}^{2}=F^{-1}dr^{2}+(r^{2}+n^{2})d\vartheta ^{2}-F(r)\left[
dt-2nw(\vartheta )d\varphi \right] ^{2}+(r^{2}+n^{2})a(\vartheta )d\varphi
^{2}  \label{pm4d}
\end{equation}%
for the so--called topological Taub--NUT--AdS/dS spacetimes \cite%
{cejm,alon,ms} with NUT charge $n.$ There are three possibilities:%
\begin{equation*}
F(r)=\frac{r^{4}+(\varepsilon l^{2}+n^{2})r^{2}-2\mu rl^{2}+\varepsilon
n^{2}(l^{2}-3n^{2})+(1-|\varepsilon |)n^{2}}{l^{2}(n^{2}+r^{2})}
\end{equation*}%
for $\varepsilon =1,0,-1$ defining respectively%
\begin{equation*}
\left\{
\begin{array}{c}
U(1)\mbox{ fibrations over }S^{2}; \\
U(1)\mbox{ fibrations over }T^{2}; \\
U(1)\mbox{ fibrations over }H^{2};%
\end{array}%
\right. \mbox{\  for \  }\left\{
\begin{array}{c}
a(\vartheta )=\sin ^{2}\vartheta , w(\vartheta )=\cos \vartheta , \\
a(\vartheta )=1, w(\vartheta )=\vartheta , \\
a(\vartheta )=\sinh ^{2}\vartheta , w(\vartheta )=\cosh \vartheta .%
\end{array}%
\right. \
\end{equation*}%
The ansatz (\ref{pm4d}) for $\varepsilon =1,0,-1$ but $n=0$ recovers
correspondingly the spherical, toroidal and hyperbolic Schwarzschild--AdS/dS
solutions of 4D Einstein equations with cosmological constant $\lambda
=-3/l^{2}$ and mass parameter $\mu .$ For our further purposes, it is
convenient to consider a coordinate transform
\begin{equation*}
(r,\vartheta ,t,\varphi )\rightarrow (r,\vartheta ,p(\vartheta ,t,\varphi
),\varphi )
\end{equation*}%
with a new time like coordinate $p$ when
\begin{equation*}
dt-2nw(\vartheta )d\varphi =dp-2nw(\vartheta )d\vartheta .
\end{equation*}%
and $t\rightarrow p$ are substituted in (\ref{pm4d}). This is possible if
\begin{equation*}
t\rightarrow p=t-\int \nu ^{-1}(\vartheta ,\varphi )d\xi (\vartheta ,\varphi
)
\end{equation*}%
with
\begin{equation*}
d\xi =-\nu (\vartheta ,\varphi )d(p-t)=\partial _{\vartheta }\xi \
d\vartheta +\partial _{\varphi }\xi  d\varphi
\end{equation*}%
when
\begin{equation*}
d(p-t)=2n w(\vartheta ) (d\vartheta -d\varphi ).
\end{equation*}%
The last formulas state that the functions $\nu (\vartheta ,\varphi )$ and $%
\xi (\vartheta ,\varphi )$ should be taken to solve the equations%
\begin{equation*}
\partial _{\vartheta }\xi =-2n w(\vartheta ) \nu \mbox{\ and \ }\partial
_{\varphi }\xi =2n w(\vartheta ) \nu .
\end{equation*}%
The solutions of such equations can be generated by any
\begin{equation*}
\xi =e^{f(\varphi -\vartheta )}\mbox{\ and \ }\nu =\frac{1}{2n w(\vartheta )%
}\frac{df}{dx}e^{f(\varphi -\vartheta )}
\end{equation*}%
for $x=\varphi -\vartheta .$

The primary ansatz (\ref{pm4d}) can be written in a form similar to (\ref%
{m1a})
\begin{eqnarray}
\mathbf{\check{g}} &=&\check{g}_{1}(x^{k},v,y^{4})(dx^{1})^{2}+\check{g}%
_{2}(x^{k},v,y^{4})(dx^{2})^{2}  \label{pm4d1} \\
&&+\check{h}_{3}(x^{k},v,y^{4})(\check{b}^{3})^{2}+\check{h}%
_{4}(x^{k},v,y^{4})(\check{b}^{4})^{2},  \notag \\
\check{b}^{3} &=&dv+\check{w}_{i}(x^{k},v,y^{4})\ dx^{i},\ \check{b}%
^{4}=dy^{4}+\check{n}_{i}(x^{k},v,y^{4})\ dx^{i},  \notag
\end{eqnarray}%
following the parametrizations
\begin{eqnarray*}
x^{1} &=&r,x^{2}=\vartheta ,y^{3}=v=p,y^{4}=\varphi \\
\check{g}_{1}(r) &=&F^{-1}(r), \check{g}_{2}(r)=(r^{2}+n^{2}), \\
\check{h}_{3}(r)&=&-F(r), \check{h}_{4}(r,\vartheta
)=(r^{2}+n^{2})a(\vartheta ), \\
\check{w}_{1}(\vartheta ) &=&-2n w(\vartheta ), \check{w}_{2}=0, \check{n}%
_{i}=0.
\end{eqnarray*}

An anholonomic transform $\mathbf{\check{N}\rightarrow N}$ and $\mathbf{%
\check{g}=}(\check{g},\check{h})\rightarrow \mathbf{g=}(g,h)$ can defined by
formulas of type (\ref{polf}),
\begin{eqnarray}
g_{1} &=&\eta _{1}(r,\vartheta )\check{g}_{1}(r),\ g_{2}=\eta
_{2}(r,\vartheta )\check{g}_{2}(r),  \label{pol4d} \\
h_{3} &=&\eta _{3}(r,\vartheta ,p)\check{h}_{3}(r),\ h_{4}=\eta
_{4}(r,\vartheta ,p)\check{h}_{4}(r,\vartheta ),  \notag \\
\ w_{1} &=&\eta _{1}^{3}(r,\vartheta ,p)\check{w}_{1}(\vartheta ),\
w_{2}=w_{2}(r,\vartheta ,p),  \notag \\
n_{1} &=&n_{1}(r,\vartheta ,p),\ n_{2}=n_{2}(r,\vartheta ,p),  \notag
\end{eqnarray}%
resulting in the ''target'' metric ansatz
\begin{eqnarray}
\mathbf{g} &=&g_{1}(r,\vartheta )(dr)^{2}+g_{2}(r,\vartheta )(d\vartheta
)^{2}+h_{3}(r,\vartheta ,p)(b^{3})^{2}+h_{4}(r,\vartheta ,p)(b^{4})^{2},
\notag \\
b^{3} &=&dp+w_{1}(r,\vartheta ,p)\ dr+w_{2}(r,\vartheta ,p)\ d\vartheta ,\
\label{4dans1} \\
b^{4} &=&d\varphi +n_{1}(r,\vartheta ,p)\ dr+n_{2}(r,\vartheta ,p)\
d\vartheta .  \notag
\end{eqnarray}%
Our aim is to state the coefficients when this off--diagonal metric ansatz
define solutions of the nonholonomic Ricci flow equations (\ref{ee1})--(\ref%
{rfe2}) for $\tau =p.$

We construct a family of exact solutions of the Einstein equations with
polarized cosmological constants following the same steps used for deriving
formulas (\ref{ee1a}), (\ref{afa1}), (\ref{solh5}), (\ref{fa1}) and (\ref%
{see4}). By a corresponding 2D coordinate transform $x^{\widetilde{i}%
}\rightarrow x^{\widetilde{i}}(r,\vartheta ),$ such metrics can be always
diagonalized and represented in conformally flat form,%
\begin{equation*}
g_{1}(r,\vartheta )(dr)^{2}+g_{2}(r,\vartheta )(d\vartheta )^{2}=e^{\psi (x^{%
\widetilde{i}})}\left[ \epsilon _{1}(dx^{\widetilde{1}})^{2}+\epsilon
_{2}(dx^{\widetilde{2}})^{2}\right] ,
\end{equation*}%
where the values $\epsilon _{i}=\pm 1$ depend on chosen signature and $\psi
(x^{\widetilde{i}})$ is a solution of
\begin{equation*}
\epsilon _{1}\psi ^{\bullet \bullet }+\epsilon _{2}\psi ^{^{\prime \prime
}}=2\lambda _{\lbrack h]}(x^{\widetilde{i}}).
\end{equation*}%
For other metric coefficients, one obtains the relations
\begin{equation*}
\phi (r,\vartheta ,p)=\ln \left| h_{4}^{\ast }/\sqrt{|h_{3}h_{4}|}\right| ,
\end{equation*}%
for
\begin{eqnarray*}
(e^{\phi })^{\ast } &=&-2\lambda _{\lbrack v]}(r,\vartheta ,p) \sqrt{%
|h_{3}h_{4}|}, \\
|h_{3}| &=&4e^{-2\phi (r,\vartheta ,p)}\left[ \left( \sqrt{|h_{4}|}\right)
^{\ast }\right] ^{2}, |h_{4}^{\ast }|=-(e^{\phi })^{\ast }/4\lambda
_{\lbrack v]}.
\end{eqnarray*}%
It is convenient to represent such solutions in the form, see(\ref{solh5}),
\begin{equation}
h_{4}=\epsilon _{4}\left[ b(r,\vartheta ,p)-b_{0}(r,\vartheta )\right]
^{2},\ h_{3}=4\epsilon _{3}e^{-2\phi (r,\vartheta ,p)}\left[ b^{\ast
}(r,\vartheta ,p)\right] ^{2}  \label{aux41}
\end{equation}%
where $\epsilon _{a}=\pm 1$ depend on fixed signature, $b_{0}(r,\vartheta )$
and $\phi (r,\vartheta ,p)$ can be arbitrary functions and $b(r,\vartheta
,p) $ is any function when $b^{\ast }$ is related to $\phi $ and $\lambda
_{\lbrack v]}$ as stated by the formula (\ref{fa1}).

The N--connection coefficients are of type
\begin{equation*}
n_{k}=n_{k[1]}(r,\vartheta )+n_{k[2]}(r,\vartheta ) \hat{n}_{k}(r,\vartheta
,p),
\end{equation*}%
where
\begin{equation*}
\hat{n}_{k}(r,\vartheta ,p)=\int h_{3}(\sqrt{|h_{4}|})^{-3}dp,\
\end{equation*}%
and $n_{k[1]}(r,\vartheta )$ and $n_{k[2]}(r,\vartheta )$ are integration
functions and $h_{4}^{\ast }\neq 0.$

The above constructed coefficients for the metric and N--connection depend
on arbitrary integration functions. One have to constrain such integral
varieties in order to construct Ricci flow solutions. Let us consider
possible solutions of the equation (\ref{rfe1}) for $n_{i}=0$ as a possible
solution of (\ref{aux31b}) (necessary for the Levi--Civita configurations).
One obtains a matrix equation for matrices $\widetilde{g}(r,\vartheta )=%
\left[ 2\lambda _{\lbrack h]}(r,\vartheta )\ g_{ij}(r,\vartheta )\right] $ $%
\ $and $\widetilde{w}(r,\vartheta ,p)=\left[ w_{i}(r,\vartheta ,p)\
w_{j}(r,\vartheta ,p)\right] $
\begin{equation}
\widetilde{g}(r,\vartheta )=h_{3}(r,\vartheta ,p)\frac{\partial }{\partial p}%
\widetilde{w}(r,\vartheta ,p).  \label{rfe1a}
\end{equation}%
This equation can be compatible for such 2D systems of coordinates when $%
\widetilde{g}$ is not diagonal (when it is more easy to contract a solution (%
\ref{ee1a}) for the diagonalized case) because $\widetilde{w}$ is also not
diagonal. For 2D subspaces the coordinate and frame transforms are
equivalent but such configurations should be corresponding adapted to the
nonholonomic structure defined by $\widetilde{w}(r,\vartheta ,p)$ which is
possible for a general 2D coordinate system. We can consider the transforms
\begin{equation*}
g_{ij}=e_{i}^{ i^{\prime }}(x^{k^{\prime }}(r,\vartheta )) e_{j}^{\
j^{\prime }}(x^{k^{\prime }}(r,\vartheta )) g_{i^{\prime }j^{\prime
}}(x^{k^{\prime }})
\end{equation*}%
and%
\begin{equation*}
w_{i^{\prime }}(x^{k^{\prime }})=e_{ i^{\prime }}^{i}(x^{k^{\prime
}}(r,\vartheta ))w_{i}((r,\vartheta ,p))
\end{equation*}%
associated to a coordinate transform $(r,\vartheta )\rightarrow x^{k^{\prime
}}(r,\vartheta ))$ with $g_{i^{\prime }j^{\prime }}(x^{k^{\prime }})$
defining, in general, a symmetric but non--diagonal ($2\times 2)$%
--dimensional matrix. We can integrate the equation (\ref{rfe1a}) in
explicit form by separation of variables in $\phi ,b,$ $h_{3}$ and $%
w_{i^{\prime }},$ when
\begin{eqnarray*}
\phi &=&\widehat{\phi }(x^{i^{\prime }}) \check{\phi}(p), h_{3}=\widehat{h}%
_{3}(x^{i^{\prime }}) \check{h}_{3}(p), \\
w_{i^{\prime }} &=&\widehat{w}_{i^{\prime }}(x^{k^{\prime }})q(p),%
\mbox{\
for }\widehat{w}_{i^{\prime }}=-\partial _{i^{\prime }}\ln |\widehat{\phi }%
(x^{k^{\prime }})|,q=(\partial _{p}\check{\phi}(p))^{-1}
\end{eqnarray*}%
where separation of variables for $\ h_{3}$ is related to a similar
separation of variables $b=\widehat{b}(x^{i^{\prime }})\check{b}(p)$ as
follows from (\ref{aux41}). One obtains the matrix equation%
\begin{equation*}
\widetilde{g}(x^{k^{\prime }})=\alpha _{0} \widehat{h}_{3}(x^{i^{\prime }})%
\widetilde{w}_{0}(x^{k^{\prime }})
\end{equation*}%
where the matrix $\widetilde{w}_{0}$ has components $\left( \widehat{w}%
_{i^{\prime }}\widehat{w}_{k^{\prime }}\right) $ and constant $\alpha
_{0}\neq 0$ is chosen from any prescribed relation%
\begin{equation}
\check{h}_{3}(p)=\alpha _{0}\partial _{p}\left[ \partial _{p}\check{\phi}(p)%
\right] ^{-2}.  \label{aux42}
\end{equation}%
We conclude that any given functions $\widehat{\phi }(x^{k^{\prime }}),%
\check{\phi}(p)$ and $\widehat{h}_{3}(x^{i^{\prime }})$ and constant $\alpha
_{0}$ we can generate solutions of the Ricci flow equation (\ref{rfe1}) for
$n_{i}=0$ with the metric coefficients parametrized in the same form as for
the solution of the Einstein equations (\ref{ee1})--(\ref{ee4}). In a
particular case, we can take $\check{\phi}(p)$ to be a periodic or solitonic
type function.

The last step in constructing flow solutions is to solve the equation (\ref%
{rfe2}) for the ansatz (\ref{4dans1}) redefined for coordinates $%
x^{k^{\prime }}=x^{k^{\prime }}(r,\vartheta ),$%
\begin{equation*}
\frac{\partial }{\partial p}h_{a}=2\lambda _{\lbrack v]}(x^{k^{\prime }},p)\
h_{a}.
\end{equation*}%
This equation is compatible if $h_{4}=\varsigma (x^{k^{\prime }})h_{3}$ for
any prescribed function $\varsigma (x^{k^{\prime }}).$ We can satisfy this
condition by corresponding parametrizations of function $\phi =\widehat{\phi
}(x^{i^{\prime }})\ \check{\phi}(p)$ and/or $b=\widehat{b}(x^{i^{\prime }})%
\check{b}(p),$ see (\ref{aux41}). As a result we can compute the effective
cosmological constant for such Ricci flows,%
\begin{equation*}
\lambda _{\lbrack v]}(x^{k^{\prime }},p)=\partial _{p}\ln
|h_{3}(x^{k^{\prime }},p)|
\end{equation*}%
which for solutions of type (\ref{aux42}) is defined by a polarization
running in time,%
\begin{equation*}
\lambda _{\lbrack v]}(p)=\alpha _{0}\partial _{p}^{2}\left[ \partial _{p}%
\check{\phi}(p)\right] ^{-2}.
\end{equation*}%
In this case we can identify $\alpha _{0}$ with a cosmological constant ($%
\lambda =-3/l^{2},$ for primary Taub-NUT configurations, or any $\lambda
=\lambda _{H}^{2}/4,$ for string configurations, see formula (\ref{strcor}))
if we choose such $\check{\phi}(p)$ that $\partial _{p}^{2}\left[ \partial
_{p}\check{\phi}(p)\right]^{-2}\rightarrow 1$ for $p\rightarrow 0.$

Putting together the coefficients of metric and N--connection, one obtains
\begin{eqnarray}
\mathbf{g} &=&\alpha _{0}\ \widehat{h}_{3}(x^{i^{\prime }})\ \{\partial
_{i^{\prime }}\ln |\widehat{\phi }(x^{k^{\prime }})|\ \partial _{j^{\prime
}}\ln |\widehat{\phi }(x^{k^{\prime }})|\ dx^{i^{\prime }}dx^{j^{\prime
}}+\partial _{p}\left[ \partial _{p}\check{\phi}(p)\right] ^{-2}\times
\notag \\
&&\left[ \left[ dp-(\partial _{p}\check{\phi}(p))^{-1}\left( dx^{i^{\prime
}}\partial _{i^{\prime }}\ln |\widehat{\phi }(x^{k^{\prime }})|\right) %
\right] ^{2}+\varsigma (x^{k^{\prime }})(d\varphi )^{2}\right] \}.
\label{sol4df}
\end{eqnarray}%
This metric ansatz depends on certain type of arbitrary integration and
generation functions $\widehat{h}_{3}(x^{i^{\prime }}),\widehat{\phi }%
(x^{k^{\prime }}),\varsigma (x^{k^{\prime }})$ and $\check{\phi}(p)$ and on
a constant $\alpha _{0}$ which can be identified with the primary
cosmological constant. It was derived by considering nonholonomic
deformations of some classes of 4D Taub-NUT solutions (\ref{pm4d1}) by
considering polarizations functions (\ref{pol4d}) deforming the coefficients
of the primary metrics into the target ones for corresponding Ricci flows.
The target metric (\ref{sol4df}) model 4D Einstein spaces with
''horizontally'' polarized, $\lambda _{\lbrack h]}(x^{k^{\prime }})$ and
''vertically'' running, $\lambda _{\lbrack v]}(p),$ cosmological constant
managed by the Ricci flow solutions.

Finally, we conclude that if the primary 4D topological Taub--NUT--AdS/dS
spacetimes have the structure of $U(1)$ fibrations over 2D hypersurfaces
(sphere, torus or hyperboloid) than their nonholonomic deformations to Ricci
flow solutions with effectively polarized/running cosmological constant
defines the generalized 4D Einstein spaces as certain foliations on the
corresponding 2D hypersurfaces. This holds true if the nonholonomic structures is chosen to
be integrable and for the Levi-Civita connection. In more general cases,
with nontrivial torsion, for instance, induced from string gravity, we deal
with "nonintegrable" foliated structures, i.e. with nonholonomic
Riemann--Cartan manifolds provided with effective nonlinear connection
structure induces by off--diagonal metric terms.

\section{Outlook and Discussion}

In summary, we have developed the method of anholonomic frames in order to
construct exact solutions describing Ricci flows of 4D Taub-NUT like
metrics. Toward this end, we applied a program of study and applications to
physics which is based on the geometry of nonholonomic/ foliated spaces with
associated nonlinear connection structure induced by generic off--diagonal
metric terms. The premise of this methodology is that one was possible to
generate exact nonholonomic solutions for three, four and five dimensional
spacetimes (in brief, 3D, 4D, 5D) in the Einstein and low/extra dimension
gravity
with cosmological constant (possibly induced by some ansatz for the
antisymmetric torsion in string gravity, or other models of gravity and
effective matter field interactions) \cite{vhep2,vs,vesnc,vt,vp}. The
validity of this approach in constructing Ricci flow solutions was
substantiated by generating certain examples of Ricci flow of solitonic
pp--waves \cite{vrf}.

In this paper, we elaborated the geometric background for generalized
Einstein spaces with effectively polarized (anisotropically on some space
coordinates and running on time like coordinate) cosmological "constants"
arising naturally if we consider generic off--diagonal gravitational (vacuum
and with nontrivial matter field interactions or extra dimension
corrections) and nonholonomic frame effects. Such spacetimes are
distinguished by corresponding nontrivial nonholonomic (foliated, if the
integrability conditions are satisfied) structures (examined in different
approaches to Lagrange--Finsler geometry \cite{ma}, semi--Riemannian and
Finsler foliations \cite{bejf} and, for instance, in modern gravity and
noncommutative geometry \cite{vesnc}). In searching for physical
applications of such geometric methods, we addressed to the geometry and
physics of Taub-NUT spacetimes \cite{haw,man,atiyah,sor,gp,vis,dv} (see also
more recent developments in Refs. \cite{onemli,amr,ar,lpp,ms}) being
interested in an analysis of the Ricci flows of Taub-NUT spaces.

There are a number of mathematical results and certain applications in
modern physics related to Ricci flow geometry \cite%
{ham1,ham2,aubin,cao,per,bakas}. Nevertheless, possible applications to
gravity theories are connected to quite cumbersome approximated methods in
constructing solutions of the Ricci flow equations. Perhaps the first
attempt to construct explicit exact solutions was performed following a
linearization approach \cite{crvis} allowing to generate exact solutions for
lower dimensions. Another class of Ricci flow solutions was related to
solitonic pp--waves in five dimensional gravity \cite{vrf}. The program of
constructing exact Ricci flow solutions can be naturally extending to an
analysis of such flows of some physically valuable metrics describing exact
solutions in gravity.

The case of flows of the so--called Taub-NUT--AdS/dS
spacetimes presents a special interest. They describe a number of very
interesting physical situations with nontrivial cosmological constant.
Certain classes of nonholonomic deformations of such solutions by generic
off--diagonal metric terms and associated anholonomic frame structures
result in another classes of exact solutions of the Einstein equations
defining the so--called locally anisotropic Taub-NUT spacetimes, see a
detailed study in Refs. \cite{vt,vp}. One of the important result of those
investigations was that the occurrence of effective polarizations on space
like coordinates and running of cosmological constant can be considered in a
manner resulting in exactly integrable systems of partial equations (to
which the Einstein equations transform for very general ansatz in 3D--5D
gravity).

The main idea of this work is to prove that effectively induced
nonhomogeneous Einstein spaces may describe Ricci flows of Taub-NUT like
metrics, for certain parametrizations of polarizations of the metric
coefficients and of the cosmological constant. We considered nonholonomic
Ricci flows of 4D Taub-NUT spaces primarily defined as fibrated structures
on 2D spherical/ toroidal/ hyperbolic hypersurfaces. Such 4D off--diagonal
flows can be also effectively modeled by families of nonhomogeneous
Einstein spaces with corresponding nonholonomic frame structure (which
transform into a foliated structure for the Levi-Civita configurations).
Here, we note that the integral varieties of the Ricci flow solutions depend
on very general classes of generating functions and integration functions
depending on three, two, one variables and on integration constants (for 3D
configurations, such functions depend on two and one variables). This is a
general property of the systems of partial equations to which the Ricci flow
and/or Einstein equations reduce for very general off--diagonal metric and
non--Riemannian linear connection ansatz.

The ansatz for usual (non--deformed) Taub-NUT spaces transform the Einstein
equations into certain systems of nonlinear ordinary differential equations
on one variable. From the very beginning such solutions were constrained to
depend only on integration constants (like the mass parameter, NUT constant
which were defined from certain physical considerations). It is a very
difficult technical task to construct exact off--diagonal solutions
depending both holonomically and anholonomically on two, three variables and
defining 4--5D spacetimes. The anholonomic frame method offers a number of
such possibilities but it results in a more sophisticate conceptual problem
for possible geometrical and physical meaning/interpretations of various
classes of integration functions and constants. In the view of such
considerations, we can argue that by imposing certain constraints on classes
of integration functions we select certain new or prescribed physical
situations when the solutions are classified by new nonlinear symmetries
(noncommutative or Lie algebra generalizations to Lie algebroid
configurations) and self--consisting embedding in solitonic and/or pp--wave
backgrounds, with anisotropic polarization of constants, deformation of
horizons and so on, see detailed discussions in Refs. \cite%
{vhep2,vs,vesnc,vt,vp,vrf}. The approach appear to be promising in
constructing exact solutions for the Ricci flows of physically important
metrics and connections when the generation and integration functions are
constrained, for instance, to define Levi-Civita configurations being
derived effectively by nonhomogeneous cosmological constants.

In conclusion, we emphasize that the nonholonomic Ricci flow solutions for
Taub--NUT like metrics continue to have a number of issues when they
are viewed from the perspective of black hole flows and cosmological
solutions, generalizations to extra dimensions non--Riemann theories of
gravity. We shall also analyze nonholnomic Ricci flows in string gravity in our further works.

\vskip5pt

\textbf{Acknowledgement: } S. V. is grateful to D. Singleton, E. Gaburov and
D. Gon\c ta for former collaboration and discussions.
He thanks the Fields Institute for hosting volunteer research work. M. V.
has been supported in
part by the MEC-CEEX Program, Romania.

\setcounter{equation}{0} \renewcommand{\theequation}
{A.\arabic{equation}}
\setcounter{subsection}{0} \renewcommand{\thesubsection}
{A.\arabic{subsection}}

\section*{Appendix A\\
The Geometry of N--connections and Anholonomic Deformations}

We outline the geometry of anholonomic deformations of geometric structures
on a Riemann--Cartan manifold. The geometric constructions will be related
to Ricci flows on such spaces and possible reductions to the Einstein
spaces. For integrable frame structures and Levi-Civita connections such
spaces transform into usual (pseudo) Riemannian fibrations.\footnote{%
The geometry of fibrations is considered in details in Ref. \cite{bejf}
following a different class of linear connections on nonholonomic manifolds
not imposing the conditions that those connections are solutions of the
Einstein, or Ricci flow, equations . In this work, we develop for Ricci
flows the approach outlined for exact solutions in Einstein and string
gravity, for instance, in Ref. \cite{vesnc}.}

\subsection{Nonholonomic transforms of vielbeins and metrics}

We consider a $\left( n+m\right) $--dimensional manifold (spacetime)\textbf{%
}$\mathbf{V},$ $n\geq 2,m\geq 1,$ \footnote{%
In this paper we shall consider four dimensional constructions and possible
reductions to three dimensions, when $n=2\,$ and $m=2,$ or $m=1$.} of
arbitrary signature enabled with a ''primary'' metric structure $\mathbf{%
\check{g}}=\check{g}\oplus _{N}\ \check{h}$ distinguished in the form
\begin{eqnarray}
\mathbf{\check{g}} &=&\check{g}_{i}(u)(dx^{i})^{2}+\check{h}_{a}(u)(\check{b}%
^{a})^{2},  \label{m1} \\
\check{b}^{a} &=&dy^{a}+\check{N}_{i}^{a}(u)dx^{i}.  \notag
\end{eqnarray}%
The local coordinates are parametrized $u=(x,y)=\{u^{\alpha
}=(x^{i},y^{a})\},$ for the indices of type $i,j,k,...=1,2,...,n$ (in brief,
horizontal, or h--indices/ components) and $a,b,c,...=n+1,n+2,...n+m$
(vertical, or v--indices/ components).\footnote{%
The Einstein's summation rule on 'up--low' indices will be applied if the
contrary case will be not emphasized.} The off--diagonal coefficients $%
\check{N}_{i}^{a}(u)$ in (\ref{m1}) state, in general, a nonintegrable $%
\left( n+m\right) $--splitting $\oplus _{N}$ in any point $u\in \mathbf{V}$
and define a class of 'N--adapted' local bases (frames, equivalently,
vielbeins) $\mathbf{\check{e}=}(\check{e},e),$ when
\begin{equation}
\mathbf{\check{e}}_{\alpha }=\left( \check{e}_{i}=\frac{\partial }{\partial
x^{i}}-\check{N}_{i}^{a}(u)\ \frac{\partial }{\partial y^{a}},e_{a}=\frac{%
\partial }{\partial y^{a}}\right) ,  \label{b1}
\end{equation}%
and local dual bases (co--frames) $\mathbf{\check{b}}=(b,\check{b}),$ when
\begin{equation}
\mathbf{\check{b}}^{\alpha }=\left( b^{j}=dx^{i},\check{b}^{b}=dy^{b}+%
\check{N}_{i}^{b}(u)\ dx^{i}\right) ,  \label{cb1}
\end{equation}%
for $\mathbf{\check{b}\rfloor \check{e}=I,}$ i.e. $\mathbf{\check{e}}%
_{\alpha }\mathbf{\rfloor }$ $\mathbf{\check{b}}^{\beta }=\delta _{\alpha
}^{\beta },$ where the inner product is denoted by '$\mathbf{\rfloor }$' and
the Kronecker symbol is denoted by $\delta _{\alpha }^{\beta }.$ The
nonintegrability of the frame structure and corresponding h- and
v--splitting in (\ref{b1}) results in the nonholonomy (equivalently,
anholonomy) relations
\begin{equation*}
\mathbf{\check{e}}_{\alpha }\mathbf{\check{e}}_{\beta }-\mathbf{\check{e}}%
_{\beta }\mathbf{\check{e}}_{\alpha }=\mathbf{\check{w}}_{\alpha \beta
}^{\gamma }\mathbf{\check{e}}_{\gamma }
\end{equation*}%
with nontrivial anholonomy coefficients
\begin{eqnarray}
\mathbf{\check{w}}_{ji}^{a} &=&-\mathbf{\check{w}}_{ij}^{a}=\mathbf{\check{%
\Omega}}_{ij}^{a}\doteqdot \check{e}_{j}\left( \check{N}_{i}^{a}\right) -%
\check{e}_{i}\left( \check{N}_{j}^{a}\right) ,  \label{anhc} \\
\mathbf{\check{w}}_{ia}^{b} &=&-\mathbf{\check{w}}_{ai}^{b}=e_{a}(\check{N}%
_{j}^{b}).  \notag
\end{eqnarray}

A metric $\mathbf{g}=g\oplus _{N}h$ parametrized in the form
\begin{eqnarray}
\mathbf{g} &=&g_{i}(u)(b^{i})^{2}+g_{a}(u)(b^{a}),  \label{m2} \\
b^{a} &=&dy^{a}+N_{i}^{a}(u)dx^{i}  \notag
\end{eqnarray}%
is a nonhlonomic transform (deformation), preserving the $(n+m)$--splitting,
of the metric $\mathbf{\check{g}}=\check{g}\oplus _{N}\ \check{h}$ if the
coefficients of (\ref{m1}) and (\ref{m2}) are related by formulas
\begin{equation}
g_{i}=\eta _{i}(u)\ \check{g}_{i},\ h_{a}=\eta _{a}(u)\ \check{h}_{a}%
\mbox{
and }N_{i}^{a}=\eta _{i}^{a}(u)\check{N}_{i}^{a},  \label{polf}
\end{equation}%
(the summation rule is not considered for the indices of 'polarizations' $%
\eta _{\alpha }=(\eta _{i},\eta _{a})$  and $\eta _{i}^{a}$ in (\ref{polf}%
)). Under anholonomic deformations, for nontrivial values $\eta _{i}^{a}(u),$
the nonholonomic frames (\ref{b1}) and (\ref{cb1}) transform correspondingly
into
\begin{equation}
\mathbf{e}_{\alpha }=\left( e_{i}=\frac{\partial }{\partial x^{i}}%
-N_{i}^{a}(u)\ \frac{\partial }{\partial y^{a}},e_{a}=\frac{\partial }{%
\partial y^{a}}\right)  \label{b1a}
\end{equation}%
and
\begin{equation}
\mathbf{b}^{\alpha }=\left( b^{j}=dx^{i},b^{c}=dy^{c}+N_{i}^{c}(u)\
dx^{i}\right)  \label{cb1a}
\end{equation}%
with the anholonomy coefficients $\mathbf{w}_{\alpha \beta }^{\gamma }$
defined by $N_{i}^{a}$ substituted in formulas (\ref{anhc}). We adopt the
convention to use ''bold'' symbols for any geometric object adapted/defined
with respect to N--elongated bases and corresponding N--connection
structures.

A set of coefficients $\mathbf{\check{N}}=\left\{ \check{N}_{i}^{a}\right\} $
states a N--connection structure on a manifold $\mathbf{V}$ if it defines a
locally nonintegrable (nonholonomic) distribution $T\mathbf{V}_{\mid u}=h%
\mathbf{V}_{\mid u}\oplus _{\check{N}}v\mathbf{V}_{\mid u}$ in any point $%
u\in $ $\mathbf{V}$ which can be globalized to a Whitney sum
\begin{equation}
T\mathbf{V}=h\mathbf{V}\oplus _{\check{N}}v\mathbf{V}.  \label{ws}
\end{equation}%
We say conventionally that a N--connection decomposes the tangent space $T%
\mathbf{V}$ into certain horizontal (h), $h\mathbf{V,}$ and vertical (v), $v%
\mathbf{V,}$ subspaces. With respect to a 'N--adapted' base (\ref{b1a}), any
vector field $\mathbf{X}$ splits into its h- and v--components,%
\begin{equation*}
\mathbf{X=}hX+vX=X^{i}\check{e}_{i}+X^{a}e_{a}
\end{equation*}%
with $X^{i}\doteqdot \mathbf{X\rfloor }b^{i}$ and $X^{a}\doteqdot \mathbf{%
X\rfloor }\check{b}^{a}.$ A similar decomposition holds for a co--vector
(1--form) $\mathbf{\tilde{X},}$
\begin{equation*}
\mathbf{\tilde{X}}=h\tilde{X}+v\tilde{X}=X_{i}b^{i}+X_{a}\check{b}^{a}.
\end{equation*}%
It should be noted that the 'interior product' '$\mathbf{\rfloor }$' is
defined by the metric structure but the 'h- and v--splitting' are stated by
the N--connection coefficients $N_{i}^{a},$ which in this work are related
to generic off--diagonal metric coefficients defined with respect to a usual
coordinate basis. \footnote{%
For simplicity, we shall omit (inverse) hats on symbols if this does not
result in confusion.}

The N--connection curvature $\mathbf{\Omega }$ of a N--connection $\mathbf{N}
$ is by definition just the Neijenhuis tensor
\begin{equation*}
\mathbf{\Omega (X,Y)}\doteqdot \left[ vX,vY\right] +v\left[ \mathbf{X,Y}%
\right] -v\left[ vX,\mathbf{Y}\right] -v\left[ \mathbf{X,}vY\right]
\end{equation*}%
where, for instance, $\left[ \mathbf{X,Y}\right] $ denotes the commutator of
vector fields $\mathbf{X}$ and $\mathbf{Y}$ on $T\mathbf{V.}$ The
coefficients $\mathbf{\check{\Omega}}_{ij}^{a}$ of a 'N--curvature' $\mathbf{%
\check{\Omega}}$ stated with respect to the bases (\ref{b1}) and (\ref{cb1}%
) are computed following the first formula in (\ref{anhc}). We can
diagonalize the metric (\ref{m2}) by certain coordinate transforms if all $%
\mathbf{w}_{\alpha \beta }^{\gamma }$ vanish, i.e. the N--connection
structure became trivial (integrable) with $\mathbf{\Omega }=0$ and $%
e_{a}(N_{j}^{b})=0.$ The subclass of of linear connections is selected as a
particular case by parametrizations of type $%
N_{a}^{b}(x,y)=A_{ai}^{b}(x^{k})y^{a}$ (for instance, in the Kaluza--Klein
gravity, the values $A_{ai}^{b}(x^{k})$ are associated to the gauge fields
after extra dimension compactifications on $y^{a}).$

An anholonomic transform $\mathbf{\check{N}\rightarrow N}$ and $\mathbf{%
\check{g}=}(\check{g},\check{h})\rightarrow \mathbf{g=}(g,h),$ defined by
formulas (\ref{polf}), deforms correspondingly the nonholonomic frame (\ref%
{b1}) and metric structures (\ref{m1}). In this paper we consider maps
(transforms) of spaces (manifolds) provided with nonlinear connection
(N--connection) structure \cite{vesnc,ma} when invariant conventional $h$--
and $v$--splitting. A manifold $\mathbf{V}$ is called N--anholonomic if it
is provided with a preferred anholonomic frame structure induced by the
generic off--diagonal coefficients of a metric \cite{vhep2,vs,vesnc}.

\subsection{Torsions and curvatures of d--connections}

A linear connection (1--form) $\mathbf{\Gamma }_{\ \alpha }^{\gamma }=%
\mathbf{\Gamma }_{\ \alpha \beta }^{\gamma }\ \mathbf{b}^{\beta }$ on $%
\mathbf{V}$ defines an operator of covariant derivation,
\begin{equation*}
\mathbf{D}=\{\mathbf{D}_{\alpha }=\left( D_{i},D_{a}\right) \}.
\end{equation*}
The coefficients
\begin{equation*}
\mathbf{\Gamma }_{ \alpha \beta }^{\gamma }\doteqdot (\mathbf{D}_{\alpha }%
\mathbf{e}_{\beta })\rfloor \mathbf{b}^{\gamma }=\left(
L_{jk}^{i},L_{bk}^{a},C_{jc}^{i},C_{bc}^{a}\right)
\end{equation*}%
can be computed in N--adapted form (i.e. with h-- and v--splitting) with
respect to the local bases (\ref{b1}) and (\ref{cb1}) following the formulas
\begin{eqnarray*}
L_{jk}^{i} &\doteqdot &\left( D_{k}e_{j}\right) \rfloor b^{i},\
L_{bk}^{a}\doteqdot \left( D_{k}e_{b}\right) \rfloor b^{a}, \\
C_{jc}^{i} &\doteqdot &\left( D_{c}e_{j}\right) \rfloor b^{i},\
C_{bc}^{a}\doteqdot \left( D_{c}e_{b}\right) \rfloor b^{a}.
\end{eqnarray*}%
Following the terminology from \cite{ma,vesnc}, we call $\mathbf{\Gamma }_{\
\alpha \beta }^{\gamma }$ to be N--distinguished (equivalently, a
d--connection) if it preserves the $\left( n+m\right) $--splitting $\oplus
_{N},$ i.e. the decomposition%
\begin{equation*}
\mathbf{D\doteqdot X}^{\alpha }\mathbf{D}_{\alpha }=X^{i}D_{i}+X^{a}D_{a}
\end{equation*}%
holds for any vector field $\mathbf{X=}X^{i}e_{i}+X^{a}e_{a},$ under
parallel transports on $\mathbf{V.}$\footnote{%
The geometrical objects defined with respect to N--adapted bases are called
respectively d--tensors, d--vectors, d--connections.}

The torsion $\mathcal{T}^{\alpha }$ of a d--connection $\mathbf{D}$ is
defined in standard form
\begin{equation}
\mathcal{T}^{\alpha }\doteqdot \mathbf{Db}^{\alpha }=\mathbf{b}^{\alpha }+%
\mathbf{\Gamma }_{\ \alpha }^{\gamma }\wedge \mathbf{b}^{\beta }.
\label{tors}
\end{equation}%
There are five types of N--adapted components of $T_{\beta \gamma }^{\alpha
} $ computed with respect to (\ref{b1a}) and (\ref{cb1a}),
\begin{eqnarray}
T_{jk}^{i} &=&L_{jk}^{i}-L_{kj}^{i},T_{bc}^{a}=C_{bc}^{a}-C_{cb}^{a},
\label{dtors} \\
T_{ja}^{i} &=&C_{ja}^{i},~T_{bi}^{a}=\frac{\partial N_{i}^{a}}{\partial y^{a}%
}-L_{bi}^{a},~T_{ji}^{a}=\mathbf{e}_{j}N_{i}^{a}-\mathbf{e}%
_{i}N_{j}^{a}=\Omega _{ij}^{a},  \notag
\end{eqnarray}%
which for some physical models can be related to a complete antisymmetric
tensor $H_{\alpha \beta \gamma }=\mathbf{e}_{[\alpha }B_{\beta \gamma ]}$ in
string gravity \cite{string1,string2} or to certain torsion fields in
(non)commutative gauge gravity \cite{vesnc}.

The curvature
\begin{equation*}
\mathcal{R}_{~\beta }^{\alpha }\doteqdot \mathbf{D\Gamma }_{\beta
}^{\alpha }=d\mathbf{\Gamma }_{\beta }^{\alpha }-\mathbf{\Gamma }_{\beta
}^{\gamma }\wedge \mathbf{\Gamma }_{\gamma }^{\alpha },
\end{equation*}%
of a d--connection $\mathbf{D,}$ splits into six types of N--adapted
components with respect to (\ref{b1a}) and (\ref{cb1a}),
\begin{equation*}
\mathbf{R}_{~\beta \gamma \delta }^{\alpha }=\left(
R_{~hjk}^{i},R_{~bjk}^{a},P_{~hja}^{i},P_{~bja}^{c},S_{~jbc}^{i},S_{~bdc}^{a}\right) ,
\end{equation*}%
where
\begin{eqnarray}
R_{\ hjk}^{i} &=&e_{k}L_{\ hj}^{i}-e_{j}L_{\ hk}^{i}+L_{\ hj}^{m}L_{\
mk}^{i}-L_{\ hk}^{m}L_{\ mj}^{i}-C_{\ ha}^{i}\Omega _{\ kj}^{a},  \notag \\
R_{\ bjk}^{a} &=&e_{k}L_{\ bj}^{a}-e_{j}L_{\ bk}^{a}+L_{\ bj}^{c}L_{\
ck}^{a}-L_{\ bk}^{c}L_{\ cj}^{a}-C_{\ bc}^{a}\Omega _{\ kj}^{c},  \notag \\
R_{\ jka}^{i} &=&e_{a}L_{\ jk}^{i}-D_{k}C_{\ ja}^{i}+C_{\ jb}^{i}T_{\
ka}^{b},  \label{dcurv} \\
R_{\ bka}^{c} &=&e_{a}L_{\ bk}^{c}-D_{k}C_{\ ba}^{c}+C_{\ bd}^{c}T_{\
ka}^{c},  \notag \\
R_{\ jbc}^{i} &=&e_{c}C_{\ jb}^{i}-e_{b}C_{\ jc}^{i}+C_{\ jb}^{h}C_{\
hc}^{i}-C_{\ jc}^{h}C_{\ hb}^{i},  \notag \\
R_{\ bcd}^{a} &=&e_{d}C_{\ bc}^{a}-e_{c}C_{\ bd}^{a}+C_{\ bc}^{e}C_{\
ed}^{a}-C_{\ bd}^{e}C_{\ ec}^{a}.  \notag
\end{eqnarray}

Contracting respectively the components of (\ref{dcurv}), $\mathbf{R}%
_{\alpha \beta }\doteqdot \mathbf{R}_{\ \alpha \beta \tau }^{\tau },$ one
computes the h- v--components of the Ricci d--tensor (there are four
N--adapted components)
\begin{equation}
R_{ij}\doteqdot R_{\ ijk}^{k},\ \ R_{ia}\doteqdot -R_{\ ika}^{k},\
R_{ai}\doteqdot R_{\ aib}^{b},\ S_{ab}\doteqdot R_{\ abc}^{c}.
\label{dricci}
\end{equation}%
The scalar curvature is defined by contracting the Ricci d--tensor with the
inverse metric $\mathbf{g}^{\alpha \beta },$
\begin{equation}
\overleftarrow{\mathbf{R}}\doteqdot \mathbf{g}^{\alpha \beta }\mathbf{R}%
_{\alpha \beta }=g^{ij}R_{ij}+h^{ab}S_{ab}.  \label{sdccurv}
\end{equation}

There are two types of preferred linear connections uniquely determined by a
generic off--diagonal metric structure with $n+m$ splitting, see $\mathbf{g}%
=g\oplus _{N}h$ (\ref{m2}):

\begin{enumerate}
\item The Levi-Civita connection $\nabla =\{\ _{\shortmid }\Gamma _{\beta
\gamma }^{\alpha }\}$ is by definition torsionless, $~\ _{\shortmid }%
\mathcal{T}=0,$ and satisfies the metric compatibility condition$,\nabla
\mathbf{g}=0$ (we shall use a left low label ''$\mid $'' in order to
emphasize that some geometric objects are constructed just for the
Levi--Civita connection).

\item The canonical d--connection $\widehat{\mathbf{\Gamma }}_{\ \alpha
\beta }^{\gamma }=\left( \widehat{L}_{jk}^{i},\widehat{L}_{bk}^{a},\widehat{C%
}_{jc}^{i},\widehat{C}_{bc}^{a}\right) $ is also metric compatible, i. e. $%
\widehat{\mathbf{D}}\mathbf{g}=0,$ but the torsion vanishes only on h-- and
v--subspaces, i.e. $\widehat{T}_{jk}^{i}=0$ and $\widehat{T}_{bc}^{a}=0,$
for certain nontrivial values of $\widehat{T}_{ja}^{i},\widehat{T}_{bi}^{a},%
\widehat{T}_{ji}^{a}.$ In this paper we shall work only with the canonical
d--connection. For simplicity, we shall omit hats on symbols and write, for
simplicity, $L_{jk}^{i}$ instead of $\widehat{L}_{jk}^{i},$ $T_{ja}^{i}$
instead of $\widehat{T}_{ja}^{i}$ and so on...but preserve the general
symbols $\widehat{\mathbf{D}}$ and $\widehat{\mathbf{\Gamma }}_{\ \alpha
\beta }^{\gamma }$
\end{enumerate}

By a straightforward calculus with respect to N--adapted frames (\ref{b1a})
and (\ref{cb1a}), one can verify that the requested properties for $\widehat{%
\mathbf{D}}$ are satisfied if
\begin{eqnarray}
L_{jk}^{i} &=&\frac{1}{2}g^{ir}\left(
e_{k}g_{jr}+e_{j}g_{kr}-e_{r}g_{jk}\right) ,  \label{candcon} \\
L_{bk}^{a} &=&e_{b}(N_{k}^{a})+\frac{1}{2}h^{ac}\left( e_{k}h_{bc}-h_{dc}\
e_{b}N_{k}^{d}-h_{db}\ e_{c}N_{k}^{d}\right) ,  \notag \\
C_{jc}^{i} &=&\frac{1}{2}g^{ik}e_{c}g_{jk},\ \widehat{C}_{bc}^{a}=\frac{1}{2}%
h^{ad}\left( e_{c}h_{bd}+e_{c}h_{cd}-e_{d}h_{bc}\right) .  \notag
\end{eqnarray}%
We note that these formulas are computed for the components of the metric $%
\mathbf{g}=g\oplus _{N}\ h$ (\ref{m2}) but in a similar form, using symbols
with ''inverse hats'' we can compute the components $\widehat{\mathbf{D}}$
of $\mathbf{\check{g}}=\check{g}\oplus _{\check{N}}\ \check{h}$ (\ref{m1})
with respect to (\ref{b1}) and (\ref{cb1}).

The Levi-Civita linear connection $\bigtriangledown =\{\ _{\shortmid }\Gamma
_{\beta \gamma }^{\alpha }\},$ uniquely defined by the conditions $~\
_{\shortmid }\mathcal{T}=0$ and $\bigtriangledown g=0,$ is not adapted to
the distribution (\ref{ws}) and its nonholonomic deformations. Let us
parametrize the coefficients in the form
\begin{equation*}
_{\shortmid }\Gamma _{\beta \gamma }^{\alpha }=\left( _{\shortmid
}L_{jk}^{i}, _{\shortmid }L_{jk}^{a}, _{\shortmid }L_{bk}^{i},\
_{\shortmid }L_{bk}^{a}, _{\shortmid }C_{jb}^{i}, _{\shortmid
}C_{jb}^{a}, _{\shortmid }C_{bc}^{i}, _{\shortmid }C_{bc}^{a}\right) ,
\end{equation*}%
where with respect to N--adapted bases (\ref{b1a}) and (\ref{cb1a})
\begin{eqnarray*}
\bigtriangledown _{\mathbf{e}_{k}}(\mathbf{e}_{j}) &=& _{\shortmid
}L_{jk}^{i}\mathbf{e}_{i}+ _{\shortmid }L_{jk}^{a}e_{a}, \bigtriangledown
_{\mathbf{e}_{k}}(e_{b})= _{\shortmid }L_{bk}^{i}\mathbf{e}_{i}+\
_{\shortmid }L_{bk}^{a}e_{a}, \\
\bigtriangledown _{e_{b}}(\mathbf{e}_{j}) &=& _{\shortmid }C_{jb}^{i}%
\mathbf{e}_{i}+ _{\shortmid }C_{jb}^{a}e_{a}, \bigtriangledown
_{e_{c}}(e_{b})= _{\shortmid }C_{bc}^{i}\mathbf{e}_{i}+ _{\shortmid
}C_{bc}^{a}e_{a}.
\end{eqnarray*}%
A straightforward calculus\footnote{%
Such results were originally considered by R. Miron and M. Anastasiei for
vector bundles provided with N--connection and metric structures, see Ref. %
\cite{ma}. Similar proofs hold true for any nonholonomic manifold provided
with a prescribed N--connection structures.} shows that the coefficients of
the Levi-Civita connection can be expressed in the form%
\begin{eqnarray}
\ _{\shortmid }L_{jk}^{i} &=&L_{jk}^{i},\ _{\shortmid
}L_{jk}^{a}=-C_{jb}^{i}g_{ik}h^{ab}-\frac{1}{2}\Omega _{jk}^{a},
\label{lccon} \\
\ _{\shortmid }L_{bk}^{i} &=&\frac{1}{2}\Omega _{jk}^{c}h_{cb}g^{ji}-\frac{1%
}{2}(\delta _{j}^{i}\delta _{k}^{h}-g_{jk}g^{ih})C_{hb}^{j},  \notag \\
\ _{\shortmid }L_{bk}^{a} &=&L_{bk}^{a}+\frac{1}{2}(\delta _{c}^{a}\delta
_{d}^{b}+h_{cd}h^{ab})\left[ L_{bk}^{c}-e_{b}(N_{k}^{c})\right] ,  \notag \\
\ _{\shortmid }C_{kb}^{i} &=&C_{kb}^{i}+\frac{1}{2}\Omega
_{jk}^{a}h_{cb}g^{ji}+\frac{1}{2}(\delta _{j}^{i}\delta
_{k}^{h}-g_{jk}g^{ih})C_{hb}^{j},  \notag \\
\ _{\shortmid }C_{jb}^{a} &=&-\frac{1}{2}(\delta _{c}^{a}\delta
_{b}^{d}-h_{cb}h^{ad})\left[ L_{dj}^{c}-e_{d}(N_{j}^{c})\right] ,\
_{\shortmid }C_{bc}^{a}=C_{bc}^{a},  \notag \\
\ _{\shortmid }C_{ab}^{i} &=&-\frac{g^{ij}}{2}\left\{ \left[
L_{aj}^{c}-e_{a}(N_{j}^{c})\right] h_{cb}+\left[ L_{bj}^{c}-e_{b}(N_{j}^{c})%
\right] h_{ca}\right\} ,  \notag
\end{eqnarray}%
where $\Omega _{jk}^{a}$ are computed as in the firs formula in (\ref{anhc})
but for the coefficients $N_{j}^{c}.$

For our purposes, it is important to state the conditions when both the
Levi--Civita connection and the canonical d--connection may be defined by the same
set of coefficients with respect to a fixed frame of reference. Following
formulas (\ref{candcon}) and (\ref{lccon}), we conclude that one holds the
component equality $_{\shortmid }\Gamma _{\beta \gamma }^{\alpha }=\widehat{%
\mathbf{\Gamma }}_{\ \alpha \beta }^{\gamma }$ if%
\begin{equation}
\Omega _{jk}^{c}=0  \label{fols}
\end{equation}%
(there are satisfied the integrability conditions and our manifold admits a
foliation structure),
\begin{equation}
\ _{\shortmid }C_{kb}^{i}=C_{kb}^{i}=0  \label{coef0}
\end{equation}%
and
\begin{equation*}
L_{aj}^{c}-e_{a}(N_{j}^{c})=0
\end{equation*}%
which, following the second formula in (\ref{candcon}), is equivalent to%
\begin{equation}
\mathbf{e}_{k}h_{bc}-h_{dc}\ e_{b}N_{k}^{d}-h_{db}\ e_{c}N_{k}^{d}=0.
\label{cond3}
\end{equation}

We conclude this section with the remark that if the conditions (\ref{fols}%
), (\ref{coef0}) and (\ref{cond3}) hold true for the metric (\ref{m2}), the
torsion coefficients (\ref{dtors}) vanish. This results in respective
equalities of the coefficients of the Riemann, Ricci and Einstein tensors.

\subsection{Gravity on N--anholonomic manifolds and foliations}

Contracting with the inverse to the d--metric (\ref{m2}) in $\mathbf{V},$ we
can introduce the scalar curvature of a d--connection $\mathbf{D,}$
\begin{equation}
{\overleftarrow{\mathbf{R}}}\doteqdot \mathbf{g}^{\alpha \beta }\mathbf{R}%
_{\alpha \beta }\doteqdot R+S,  \label{dscal}
\end{equation}%
where $R\doteqdot g^{ij}R_{ij}$ and $S\doteqdot h^{ab}S_{ab}$ and compute
the Einstein tensor
\begin{equation}
\mathbf{G}_{\alpha \beta }\doteqdot \mathbf{R}_{\alpha \beta }-\frac{1}{2}%
\mathbf{g}_{\alpha \beta }{\overleftarrow{\mathbf{R}}.}  \label{deinst}
\end{equation}%
In the vacuum case, $\mathbf{G}_{\alpha \beta }=0,$ which mean that all
Ricci d--tensors (\ref{dricci}) vanish.

The Einstein equations for the canonical d--connection $\mathbf{\Gamma }_{\
\alpha \beta }^{\gamma }$ (\ref{candcon}),
\begin{equation}
\mathbf{R}_{\alpha \beta }-\frac{1}{2}\mathbf{g}_{\alpha \beta }%
\overleftarrow{\mathbf{R}}=\kappa \mathbf{\Upsilon }_{\alpha \beta },
\label{einsteq}
\end{equation}%
are defined for a general source of matter fields and, for instance,
possible string corrections, $\mathbf{\Upsilon }_{\alpha \beta }.$ It should
be emphasized that there is a nonholonomically induced torsion $\mathbf{T}%
_{\ \alpha \beta }^{\gamma }$ with d--torsions computed by introducing
consequently the coefficients of d--metric (\ref{m2}) into (\ref{candcon})
and than into formulas (\ref{dtors}). The gravitational field equations (\ref%
{einsteq}) can be decomposed into h-- and v--components following formulas (%
\ref{dricci}) and (\ref{dscal}),%
\begin{eqnarray}
R_{ij}-\frac{1}{2}g_{ij}\left( R+S\right) &=&\mathbf{\Upsilon }_{ij},
\label{ep1} \\
S_{ab}-\frac{1}{2}h_{ab}\left( R+S\right) &=&\mathbf{\Upsilon }_{ab},  \notag
\\
\ ^{1}P_{ai} &=&\mathbf{\Upsilon }_{ai},  \notag \\
\ -\ ^{2}P_{ia} &=&\mathbf{\Upsilon }_{ia}.  \notag
\end{eqnarray}%
The vacuum equations, in terms of the Ricci tensor $R_{\ \beta }^{\alpha
}=g^{\alpha \gamma }R_{\gamma \beta },$ are
\begin{equation}
R_{j}^{i}=0,S_{b}^{a}=0,\ ^{1}P_{i}^{a}=0,\ ^{2}P_{a}^{i}=0.  \label{ep2}
\end{equation}%
If the conditions (\ref{fols}), (\ref{coef0}) and (\ref{cond3}) are
satisfied, the equations (\ref{ep1}) and (\ref{ep2}) are equivalent to those
derived for the Levi-Civita connection. In such cases, the spacetime is
modelled as foliated manifold with generic off--diagonal metric (\ref{m2})
if the anholonomy coefficients $\mathbf{w}_{\alpha \beta }^{\gamma },$
defined by $N_{i}^{a}$ substituted in formulas (\ref{anhc}), are not zero.

In string gravity the nontrivial torsion components (\ref{dtors}) and source
$\kappa \mathbf{\Upsilon }_{\alpha \beta }$ can be related to certain
effective interactions with the strength (torsion)
\begin{equation*}
H_{\mu \nu \rho }=\mathbf{e}_{\mu }B_{\nu \rho }+\mathbf{e}_{\rho }B_{\mu
\nu }+\mathbf{e}_{\nu }B_{\rho \mu }
\end{equation*}%
of an antisymmetric field $B_{\nu \rho },$ when%
\begin{equation}
R_{\mu \nu }=-\frac{1}{4}H_{\mu }^{\ \nu \rho }H_{\nu \lambda \rho }
\label{c01}
\end{equation}%
and
\begin{equation}
D_{\lambda }H^{\lambda \mu \nu }=0,  \label{c02}
\end{equation}%
see details on string gravity, for instance, in Refs. \cite{string1,string2}%
. The conditions (\ref{c01}) and (\ref{c02}) are satisfied by the ansatz
\begin{equation}
H_{\mu \nu \rho }=\widehat{Z}_{\mu \nu \rho }+\widehat{H}_{\mu \nu \rho
}=\lambda _{\lbrack H]}\sqrt{\mid g_{\alpha \beta }\mid }\varepsilon _{\nu
\lambda \rho }  \label{ansh}
\end{equation}%
where $\varepsilon _{\nu \lambda \rho }$ is completely antisymmetric and the
distortion (from the Levi--Civita connection) and
\begin{equation*}
\widehat{Z}_{\mu \alpha \beta }\mathbf{c}^{\mu }=\mathbf{e}_{\beta }\rfloor
\mathcal{T}_{\alpha }-\mathbf{e}_{\alpha }\rfloor \mathcal{T}_{\beta }+\frac{%
1}{2}\left( \mathbf{e}_{\alpha }\rfloor \mathbf{e}_{\beta }\rfloor \mathcal{T%
}_{\gamma }\right) \mathbf{c}^{\gamma }
\end{equation*}%
is defined by the torsion tensor (\ref{tors}) with coefficients (\ref{dtors}%
). We emphasize that our $H$--field ansatz is different from those already
used in string gravity when $\widehat{H}_{\mu \nu \rho }=\lambda _{\lbrack
H]}\sqrt{\mid g_{\alpha \beta }\mid }\varepsilon _{\nu \lambda \rho }.$  In
our approach, we define $H_{\mu \nu \rho }$ and $\widehat{Z}_{\mu \nu \rho }$
from the respective ansatz for the $H$--field and nonholonomically deformed
metric, compute the torsion tensor for the canonical distinguished
connection and, finally, define the 'deformed' H--field as $\widehat{H}_{\mu
\nu \rho }=\lambda _{\lbrack H]}\sqrt{\mid g_{\alpha \beta }\mid }%
\varepsilon _{\nu \lambda \rho }-\widehat{Z}_{\mu \nu \rho }.$ Such
spacetimes are both nonholonomic with nontrivial torsion related to that in
string gravity and with sources induced by string corrections via an
effective cosmological constant $\lambda _{\lbrack H]},$ when
\begin{equation}
\mathbf{R}_{\ \beta }^{\alpha }=-\frac{\lambda _{\lbrack H]}^{2}}{4}\ {\delta}_{\ \beta }^{\alpha }.
\label{strcor}
\end{equation}%
In order to generate solutions for such equations it is more convenient to
work directly with the canonical d--connection (\ref{candcon}).

\end{document}